\documentclass{raa}  
\usepackage{graphicx,times}
\usepackage{natbib}
\usepackage{amssymb,amsmath}
\bibpunct{(}{)}{;}{a}{}{,}

\usepackage{soul,xcolor}
\setstcolor{red}

\usepackage[a4paper=true,pagebackref=true]{hyperref}
\hypersetup{pdftitle = The title of my PDF, pdfauthor = My name, pdfsubject= The subject, pdfkeywords = keyword1 keyword2 keyword3}
\hypersetup{colorlinks = true, linkcolor = green, anchorcolor = red, citecolor = blue, filecolor = red, pagecolor = red, urlcolor = red}

\begin{document}

   \title{New continuum observations of the Andromeda galaxy M 31 with FAST}

 \volnopage{Vol.0 (20xx) No.0, 000--000}
   \setcounter{page}{1}

   \author{Wenjun Zhang\inst{1}, Xiaohui Sun\inst{1}, Jie Wang\inst{2}
   }

   \institute{School of Physics and Astronomy, Yunnan University, Kunming 650500, China; {\it zhangwenjun@mail.ynu.edu.cn, xhsun@ynu.edu.cn}\\
        \and
             National Astronomical Observatories, Chinese Academy of Sciences, Beijing 100101, China\\
\vs \no
   {\small Received 20XX Month Day; accepted 20XX Month Day}
}

\abstract{We present a new total intensity image of M~31 at 1.248~GHz, observed with the Five-hundred-meter Aperture Spherical radio telescope (FAST) with an angular resolution of $4\arcmin$ and a sensitivity of about 16~mK. The new FAST image clearly reveals weak emission outside the ring due to its high sensitivity on large-scale structures. We derive a scale length of 2.7~kpc for the cosmic ray electrons and find that the cosmic ray electrons propagate mainly through diffusion by comparing the scale length at 4.8~GHz. The spectral index of the total intensity varies along the ring, which can be attributed to the variation of the spectra of synchrotron emission. This variation is likely caused by the change of star formation rates along the ring. We find that the azimuthal profile of the non-thermal emission can be interpreted by an axisymmetric large-scale magnetic field with varying pitch angle along the ring, indicating a complicated magnetic field configuration in M~31. 
\keywords{galaxies: individual: M~31 -- galaxies: ISM -- radiation mechanisms: non-thermal, thermal -- ISM: cosmic rays -- radio continuum: ISM}
}

   \authorrunning{Zhang et al.}            %author_head in even pages
   \titlerunning{M 31 with FAST}  % title_head in odd pages
   \maketitle

%________________________________________________ sections below
% 
\section{Introduction}           %% first-level sections will be auto-capitalized
\label{sect:intro}

The Andromeda galaxy (M~31) is a prominent spiral galaxy that is the nearest to the Milky Way~(\citealt{Beck+1998}). Due to its proximity and outstanding ``ring" of star formation, M~31 has been widely studied in various bands, such as X-ray (\citealt{2005A&A...434..483P}), UV (\citealt{2007ApJS..173..185G}), near-IR (\citealt{2003AJ....125..525J}), far-IR (\citealt{1988ApJS...68...91R}), and optical (\citealt{2020MNRAS.497..687S}). These observations have improved our understanding of the formation and evolution of M~31.

The radio total intensity of M~31 is concentrated in a ring about 7--13~kpc from the center (\citealt{Beck+1998}). In this ring, there are also concentrations of cold molecular gas (\citealt{2006A&A...453..459N}), warm neutral gas (\citealt{2009ApJ...705.1395C}), warm ionized gas (\citealt{1994AJ....108.1667D}), and dust (\citealt{2012ApJ...756...40S}). All these indicate that the M~31 ring is an important star-forming region and contributes most of the brightness of the galaxy. 

There have been many studies based on the radio observations of M~31. The first radio observation of M~31 was conducted with the Cambridge one-mile radio telescope at 408 and 1407~MHz with resolutions of $80\arcmin \times 120\arcmin$ and $23\arcmin \times 35\arcmin$, respectively~(\citealt{1969MNRAS.144..101P}). With the continuing improvement of angular resolution and sensitivity, the follow-up observations of M~31 reveals more detailed emission structures, such as the observations by the Effelsberg 100-m telescope at 4850~MHz with a resolution of $2\farcm6$ (\citealt{1983A&A...117..141B}). In addition, the combination of interferometric observations from the Very Large Array and the single-dish observations from the Effelsberg radio telescope at 1.5~GHz yielded an image with a high resolution of $45\arcsec$ covering all the spatial scales (\citealt{Beck+1998}).

The spectrum of M~31 has been studied based on the radio observations. The radio emission consists of synchrotron emission (non-thermal) and free-free emission (thermal). At frequencies below 10~GHz, the non-thermal emission is dominant~(\citealt{2019ApJ...877L..31B}), and the fraction of thermal emission is only about 24\% at 1~GHz~(\citealt{Berkhuijsen+2003}). Based on the observations at 1.46~GHz and 4.85~GHz, \cite{Berkhuijsen+2003} obtained an overall non-thermal spectral index of $\alpha = -1\pm 0.1$ ($S_\nu\propto\nu^\alpha$, with $S_\nu$ being the flux density at frequency $\nu$). \cite{Beck+2020} derived a total intensity spectral index of $-0.71 \pm 0.02$, and a synchrotron emission spectral index of $-0.81 \pm 0.03$ using the observations at 2.645~GHz, 4.85~GHz, and 8.35~GHz. The spectral index towards the ring increases to $-$0.5 because the thermal fraction is large and the spectrum thus tends to be flat (\citealt{2021A&A...651A..98F}). 

The Five-hundred-meter Aperture Spherical radio Telescope (FAST) is well suited to image extended sources such as M~31 with its high sensitivity~\citep{2020RAA....20...64J}. The observations of M~31 Halo is one of the key projects currently being conducted by FAST with guaranteed time (PI: Jie Wang), which focuses on HI mapping in the region of about 800 square degree around the M~31 and pulsar searching in the M~31 galaxy. The continuum and polarization data have also been recorded simultaneously. This allows us to analyse the total intensity distribution and the spatial variation of spectral index to infer the properties of M~31.

The paper is organized as the follows: the observations and results are described in Sect.~\ref{sect:Obs}, and the results are discussed in Sect.~\ref{sect:discussion} , and summarized in Sect.~\ref{sect:Summary}.

\section{Observations and Results}
\label{sect:Obs}

M~31 was observed in  drifting-scan mode (\citealt{2020RAA....20...64J}). In this mode, the 19-beam receiver is rotated by $23\fdg4$ so that a declination strip of about $22\arcmin$ can be completed in Nyquist sampling for each individual scan. In total, 11 drifting scans were conducted during October 1 -- 18, 2020. In order to calibrate the system variations, a linearly polarized noise of about 12.5~K (\citealt{2019SCPMA..6259502J}) is injected into the system for 2 seconds every 1000 seconds. The frequency range of the receiving system is 1.0 -- 1. 5~GHz, which is divided into 65536 channels. The width of each channel is about 7.63~kHz. The backend outputs four channels with two for total intensity $I_1$  and $I_2$ , and the other two for Stokes $U$ and $V$. 

We follow the data processing procedure developed by \cite{2021RAA....21..282S}. To obtain the images at a single frequency channel, the following steps were taken: (1) mitigation of radio frequency interference (RFI); (2) correction of instrumental polarization leakage and angle, and temperature scale with injected noise; (3) combination of all the 11 scans; (4) conversion to the brightness temperature scale ($T_{\rm b}$). We repeat this procedure and obtain the maps of about 29,000 frequency channels. These maps are then smoothed to a common resolution of $4\arcmin$.

\begin{figure} 
   \centering
   \includegraphics[width=0.98\textwidth, angle=0]{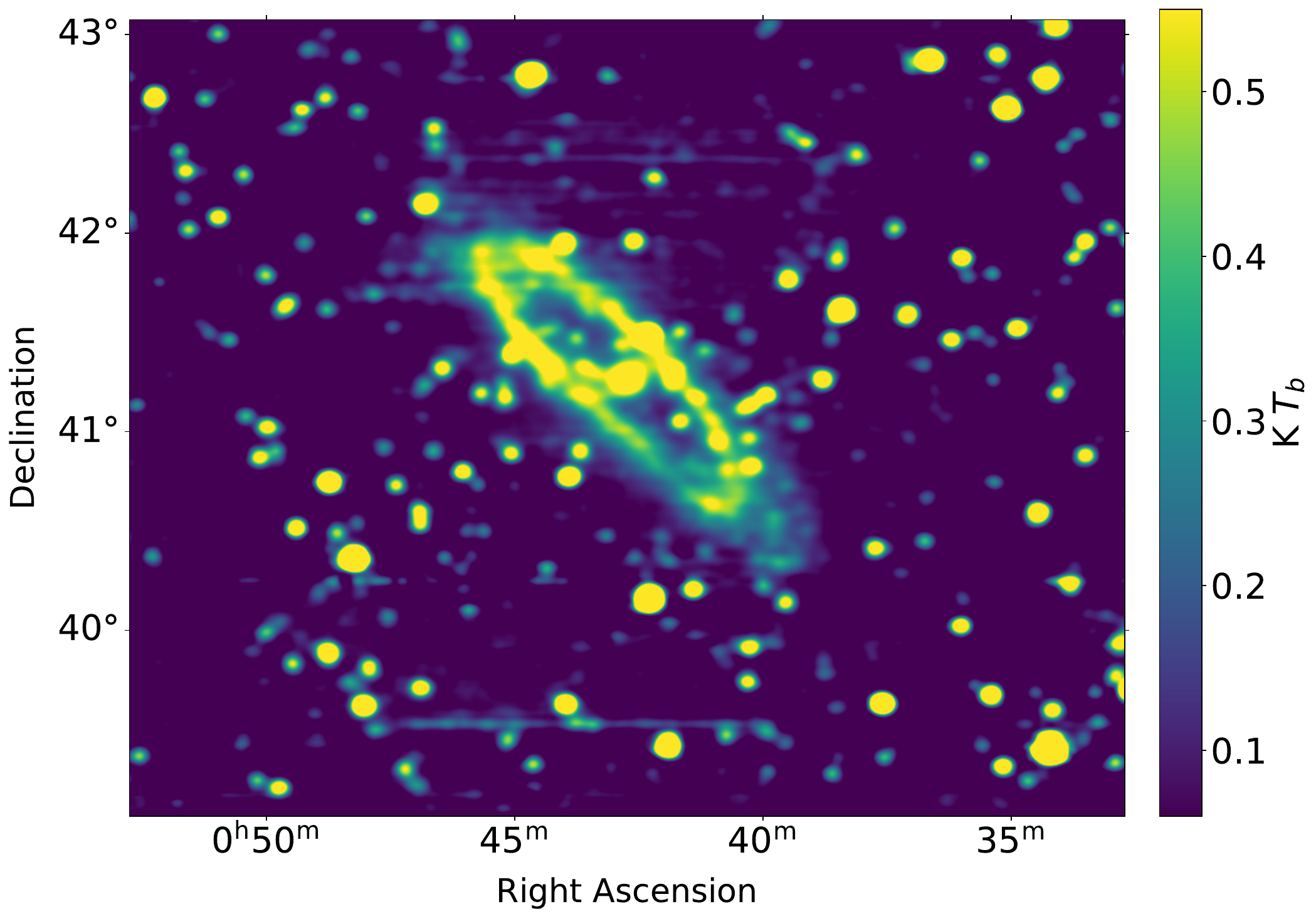}
   \caption{Total intensity ($I$) image of M 31 at 1.248~GHz from FAST. The resolution is $4\arcmin$ and the rms noise is 16~mK.} 
   \label{tot_intensity}
\end{figure}

%\section{RESULTS }
%\label{sec:RESULTS}

\subsection{Total intensity maps}

\begin{figure} 
   \centering
   \includegraphics[width=0.9\textwidth, angle=0]{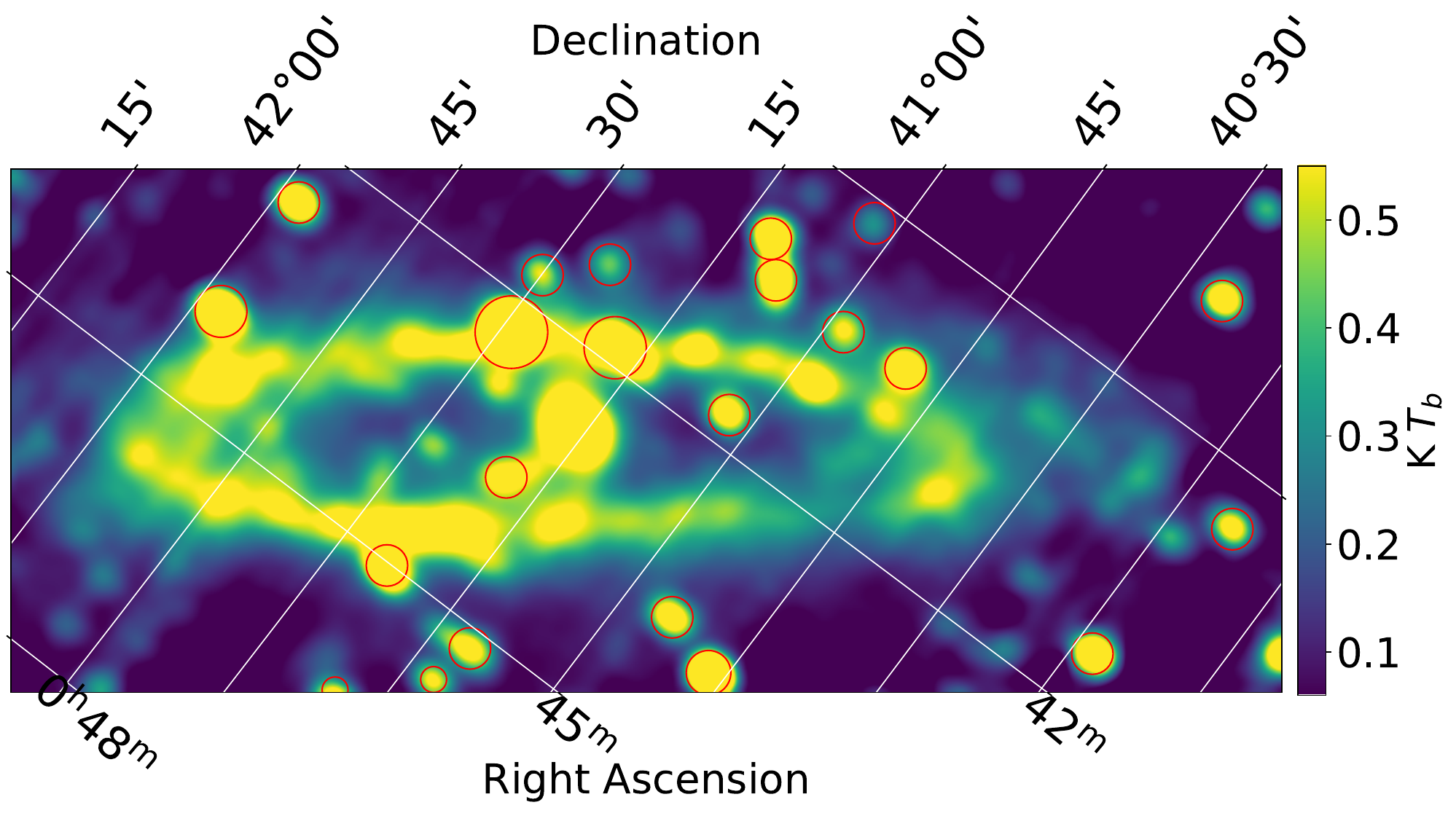}
   \includegraphics[width=0.9\textwidth, angle=0]{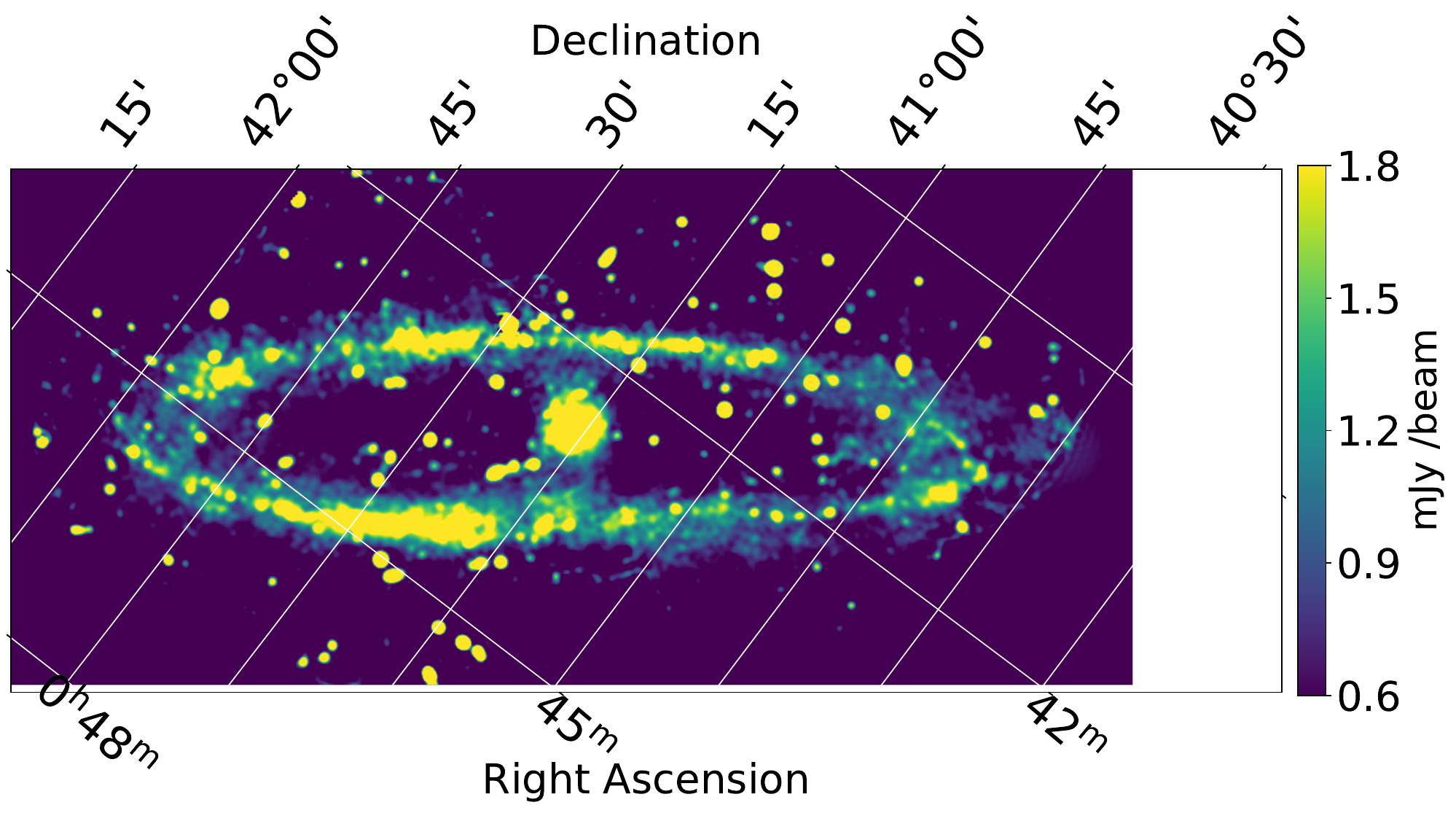}
   \includegraphics[width=0.9\textwidth, angle=0]{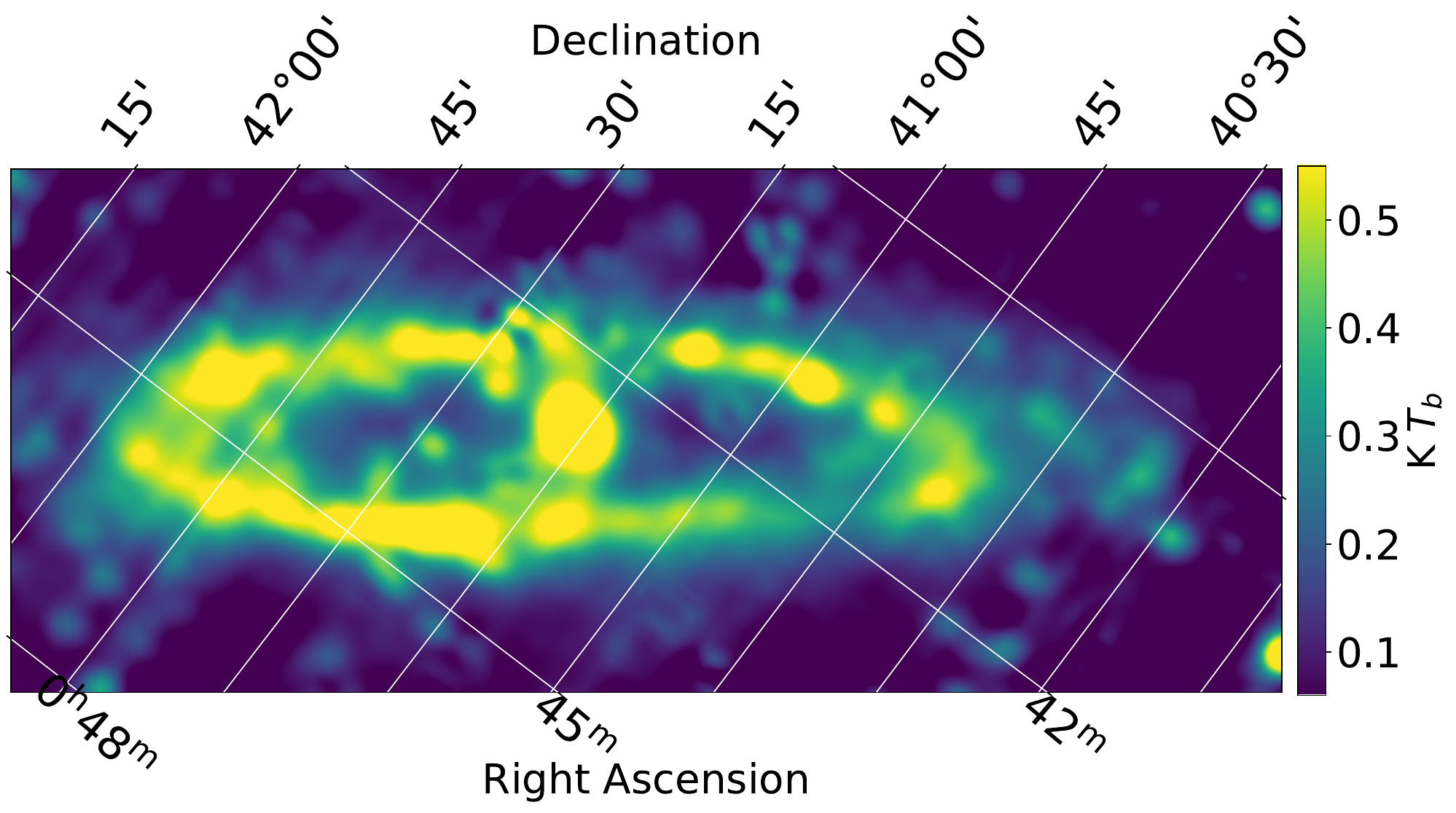}
   \caption{Total intensity ($I$) image of M~31 from FAST (top panel), from the combination of VLA and Effelsberg 100~m telescope (middle panel, \cite{Beck+1998}), and from FAST with point sources marked in the upper panel removed (bottom panel ). The galaxy is rotated by $-53\degr$.} 
   \label{tot_intensity_part}
\end{figure}
   
We average all the frequency channel maps by taking the median values and obtained the total intensity image at 1.248~GHz, shown in Fig.~\ref{tot_intensity}. Some scanning effects indicated by the stripes along declination direction still remain in the image, which were caused by drifting of the system. It would require scans in alternative directions to remove these effects, as shown by \citet{sun+22}. The rms noise is about 16~mK.

A smaller area containing M~31 is cut out from Fig.~\ref{tot_intensity}, rotated by $53\degr$ following \citet{Beck+1998} and shown in Fig.~\ref{tot_intensity_part} (top panel). For comparison, the map by \citet{Beck+1998} combining the interferometer and single-dish observations at 1.46~GHz is also shown in Fig.~\ref{tot_intensity_part} (middle panel).

As can be seen from Figs.~\ref{tot_intensity} and \ref{tot_intensity_part}, there is strong radio emission towards the center and the ring area. The FAST image clearly reveals weak emission extending outside the ring towards the southwest. In contrast, this extended emission is almost absent in the image by \cite{Beck+1998}.

To study the emission from M~31, we remove the bright compact sources marked in Fig.~\ref{tot_intensity_part} (top panel) by subtracting Gaussian fittings of these sources. The resulted image is shown in Fig.~\ref{tot_intensity_part} (bottom panel), which is used to investigate the variations of total intensity below. Note that there still remain background point sources that cannot be resolved with FAST. We retrieve the flux intensities of these sources from the NVSS survey with a higher resolution of 45$\arcsec$~\citep{Condon+1998}, and find that their contribution to the intensity of the ring is less than about 5\%. Their influence to the analyses below can thus be ignored.
%{The point sources, which are only observable at higher resolutions, were analyzed using the NVSS data at 1.4 GHz with a resolution of $45^{\prime \prime}$~\citep{Condon+1998}. Our findings indicated that the effect is negligible, with a maximum impact of 5\% on the radial profile of the total intensity and it has an insignificant impact on the azimuthal profile, remaining within the error bar range.} 

\begin{figure}  
   \centering
   \includegraphics[width=0.98\textwidth, angle=0]{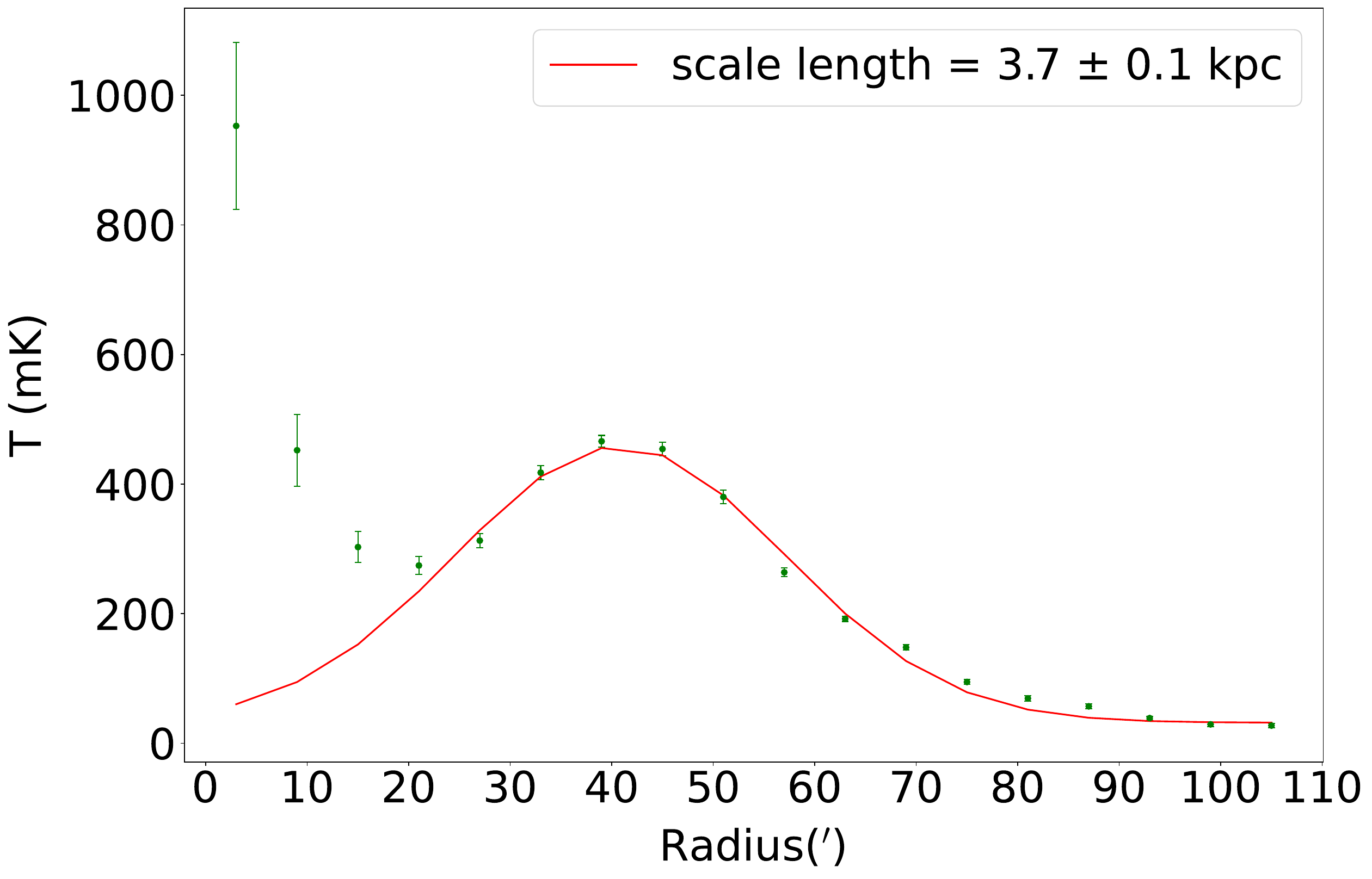}
   \caption{Intensity $I$ with the angular distance from the galactic center. The red line indicates a Gaussian fitting with a scale length of 3.7~kpc after correcting the influence of beam width.} 
   \label{intensity_trend}
   \end{figure}

\subsection{Radial profile of the total intensity}
\label{Radial distribution of intensity}
The radial variation of the total intensity can be used to constrain the propagation of cosmic ray electrons. We derive the average intensity of $6\arcmin$-width annuli from the center till the radius of about $110\arcmin$. The intensity versus radius is shown in Fig.~\ref{intensity_trend}. The standard deviation from each annulus is calculated as the error bar.

The sharp peak near radius 0 is contributed by the nucleus of M~31. The broad peak from the radius of about $27\arcmin$ to the radius of about $57\arcmin$ indicates the ring area with bright emission. With a distance to M~31 of about 780~kpc~\citep{1998ApJ...503L.131S}, this angular range corresponds to about 6 -- 13~kpc. The total intensity gradually decreases with increasing radius and flattens out beyond the radius of about $93\arcmin$. 

We fit the total intensity radial profile for radii larger than $27\arcmin$ to a Gaussian, 
%\textcolor{blue}{\bf We acquired 1000 samples of the radial profile for the total intensity, specifically for radii larger than $27\arcmin$, and determined their respective standard deviation using the bootstrap algorithm. Subsequently, we employed a Gaussian fitting approach to analyze these samples,} 
namely $T\propto \exp[-(r-r_0)^2/2L^2]$, where $r$ is the radius, $r_0$ is the peak radius, and $L$ is the scale length. After accounting for the influence of beam width, we obtain the scale length of about $16\farcm2\pm0\farcm3$ or $3.7\pm0.1$~kpc.
%\textcolor{blue}{The error in the scale length fitting was determined by calculating the standard deviation among the results obtained from multiple fits.}

\subsection{Azimuthal profile of the total intensity}
\begin{figure}  
   \centering
   \includegraphics[width=0.98\textwidth, angle=0]{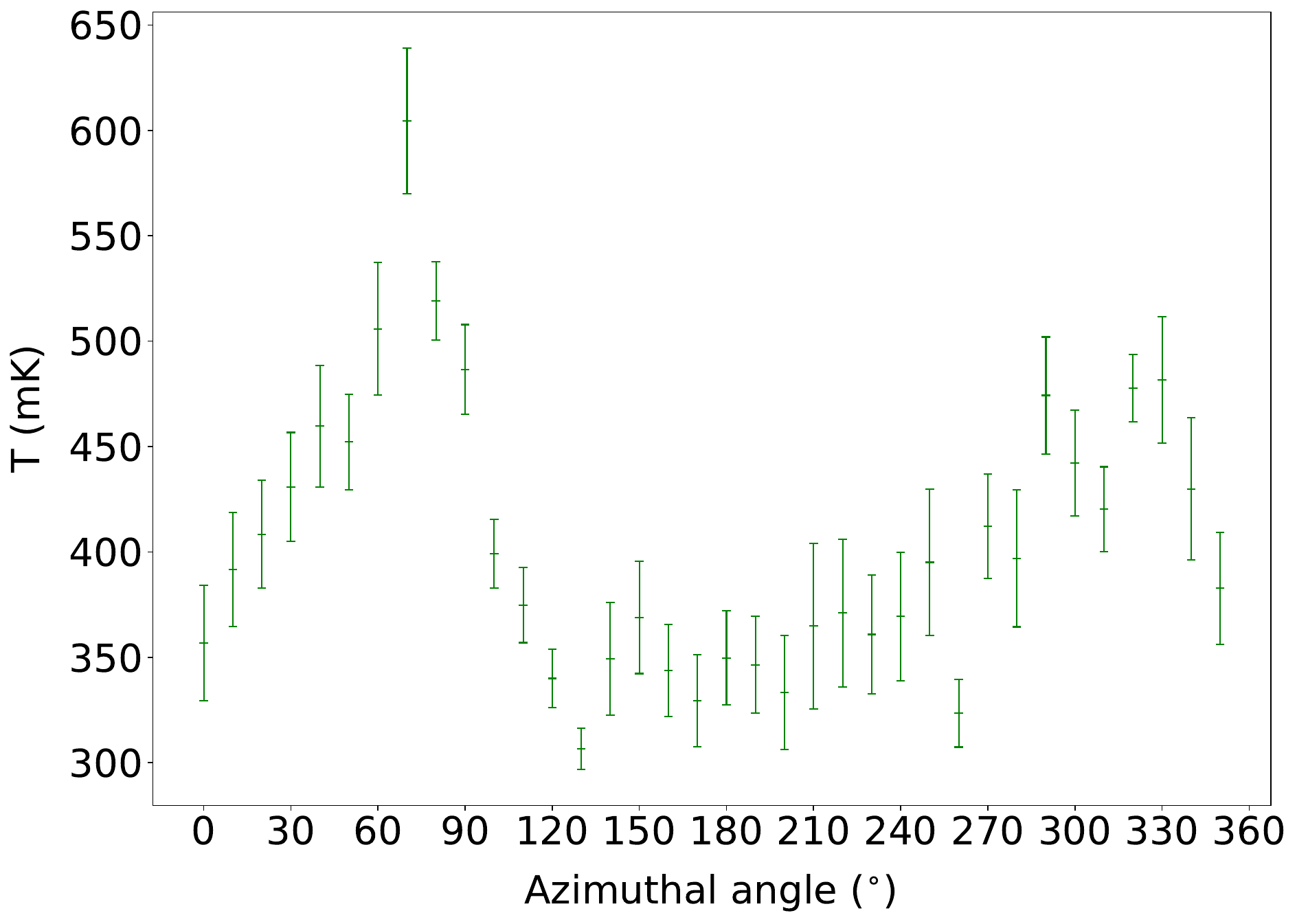}
   \caption{The total intensity in the ring of M~31 versus azimuthal angle.} 
   \label{I_Azimuthal_angle}
   \end{figure}

The variation of total intensity versus direction reflects the structure of magnetic field. From the Gaussian fitting of the radial profile, we obtain the peak position of the ring at $r_0=40\farcm9$. We use the annulus of $r_0\pm L$ to derive the azimuthal profile of total intensity in the ring. The azimuthal angle starts from the major axis towards the north east and increases eastwards (Fig.~\ref{tot_intensity}) or starts from the major axis towards the left and increases counter-clockwise (Fig.~\ref{tot_intensity_part}). An average intensity is derived for every $10\degr$ sector, and the profile is shown in Fig.~\ref{I_Azimuthal_angle}.

The emission is the brightest for the azimuthal angle between $30\degr$ and $90\degr$, and becomes weaker for the azimuthal angle between $150\degr$ and $270\degr$, as can also be seen from Fig.~\ref{tot_intensity_part}. There is no clear pattern for the azimuthal profile.

\subsection{Spectral index}
\begin{figure}
   \centering
   \includegraphics[width=0.98\textwidth, angle=0]{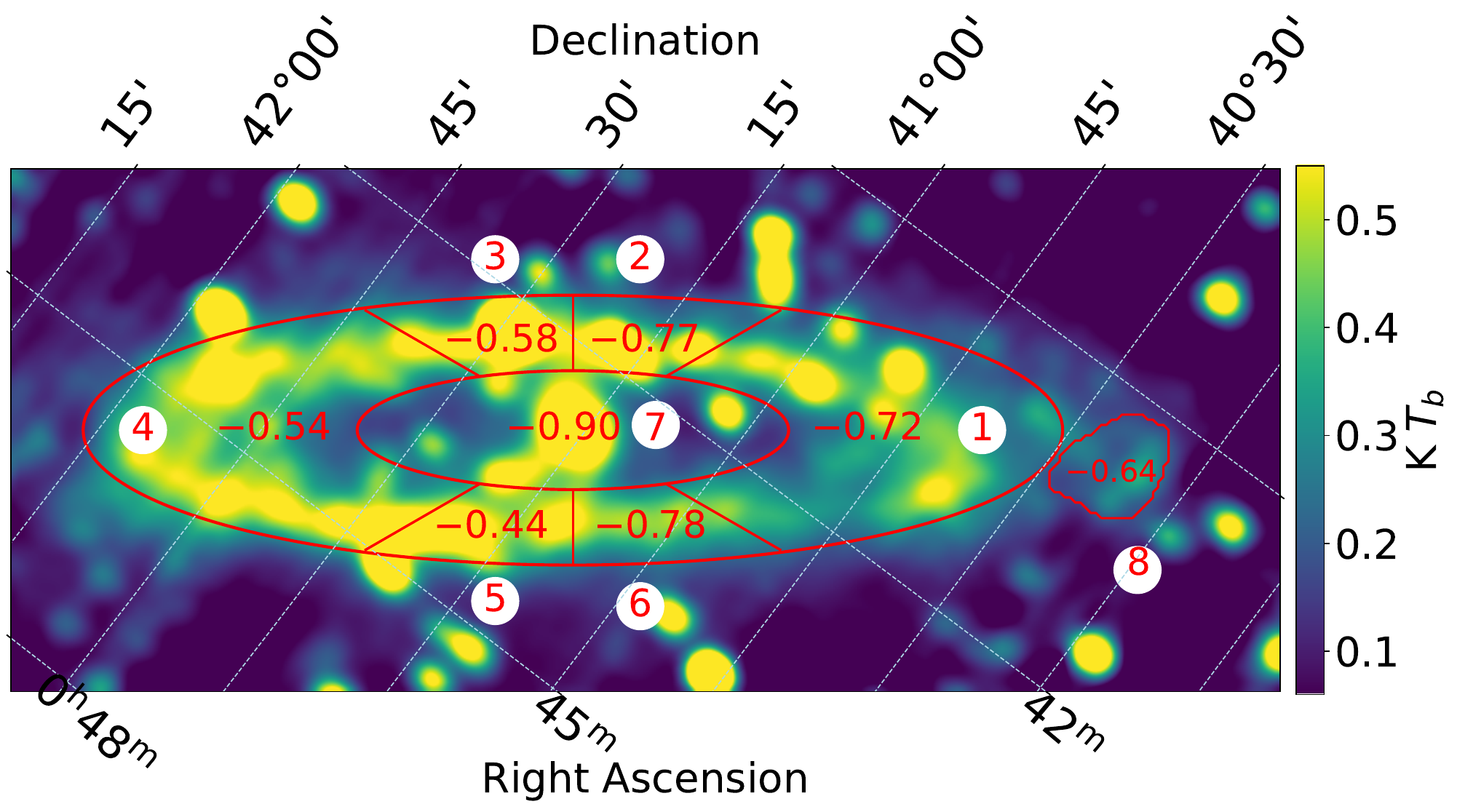}
   \caption{Area marked for deriving spectral index with TT-plot method and the resulted flux density spectral indices $\alpha=\beta+2$.}
   \label{T-T_picture_mark}
   \end{figure}

The spectral index is related with the emission properties, such as non-thermal and thermal fraction and cosmic ray electrons aging. We do not derive the spectral index for all the pixels by fitting the brightness temperature against frequency because of the influence of background emission. Instead, we divide the whole M~31 area into 8 regions~(Fig.~\ref{T-T_picture_mark}) and obtain the spectral index using Temperature-Temperature plot (TT-plot) method~\citep{turtle+62} which is immune to the background emission.

\begin{figure}[htbp]
	\centering
	\begin{subfigure}{0.49\linewidth}
		\centering
		\includegraphics[width=1\linewidth]{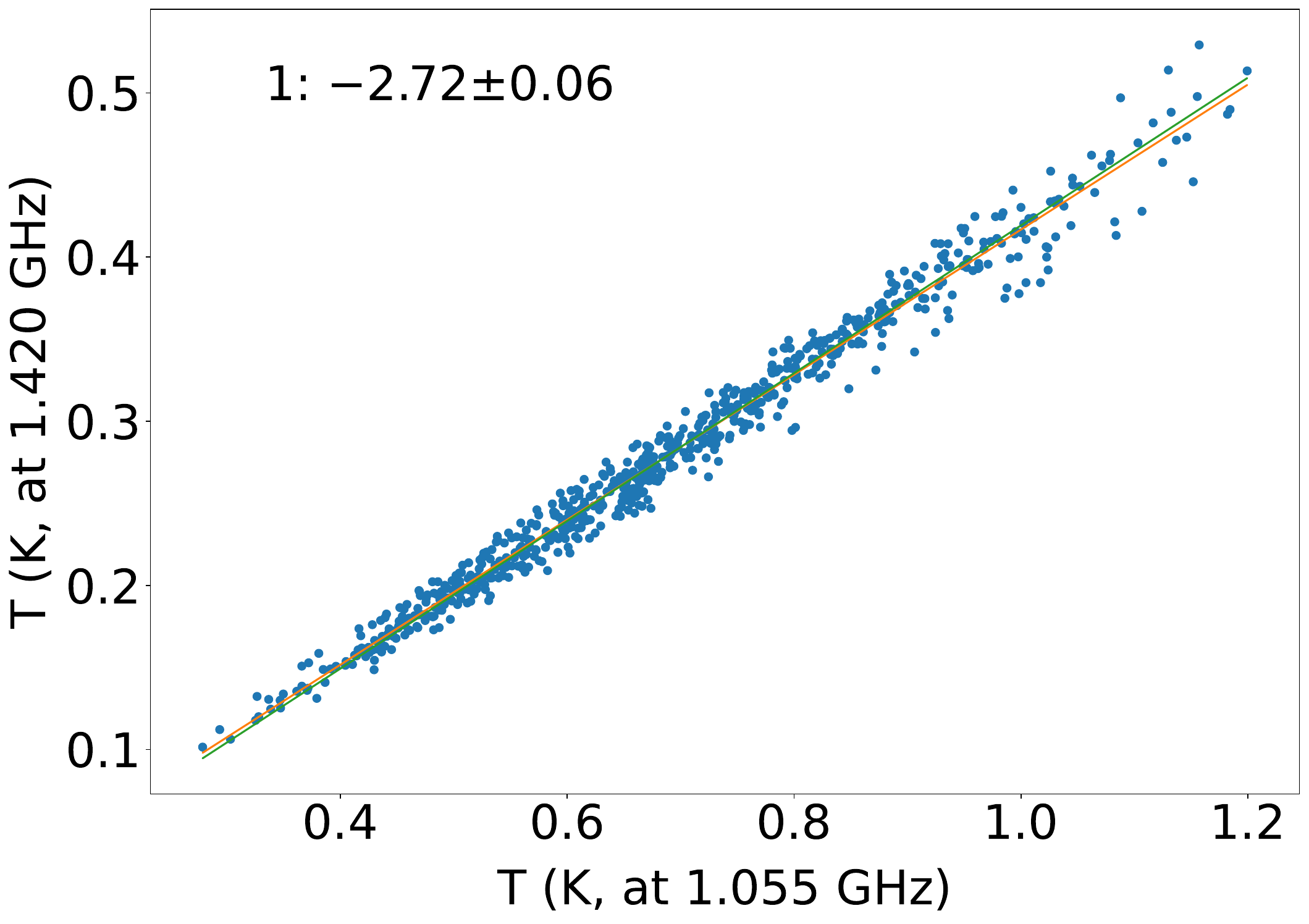}
	\end{subfigure}
	\centering
	\begin{subfigure}{0.49\linewidth}
		\centering
		\includegraphics[width=1\linewidth]{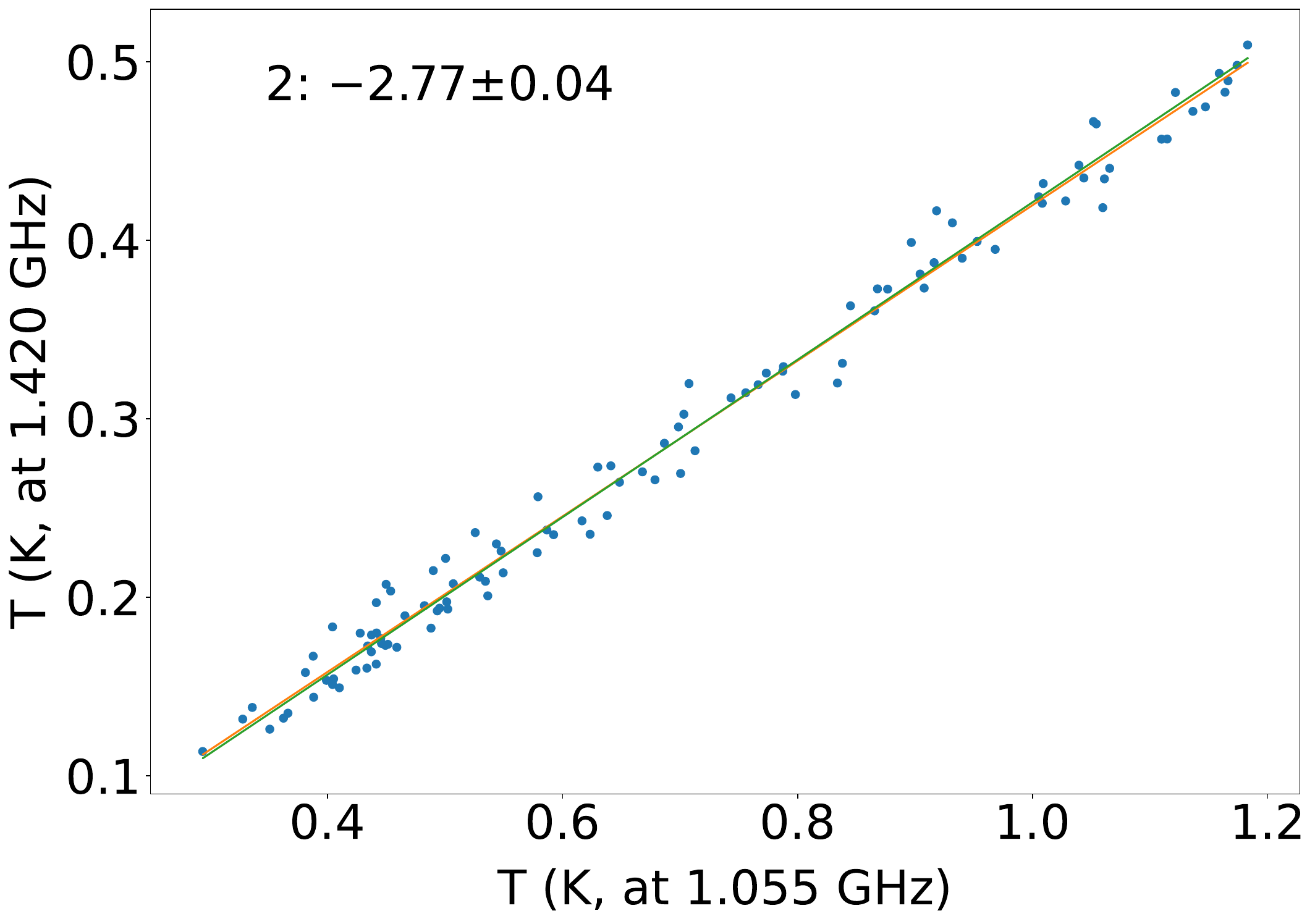}
	\end{subfigure}

	\begin{subfigure}{0.49\linewidth}
		\centering
		\includegraphics[width=1\linewidth]{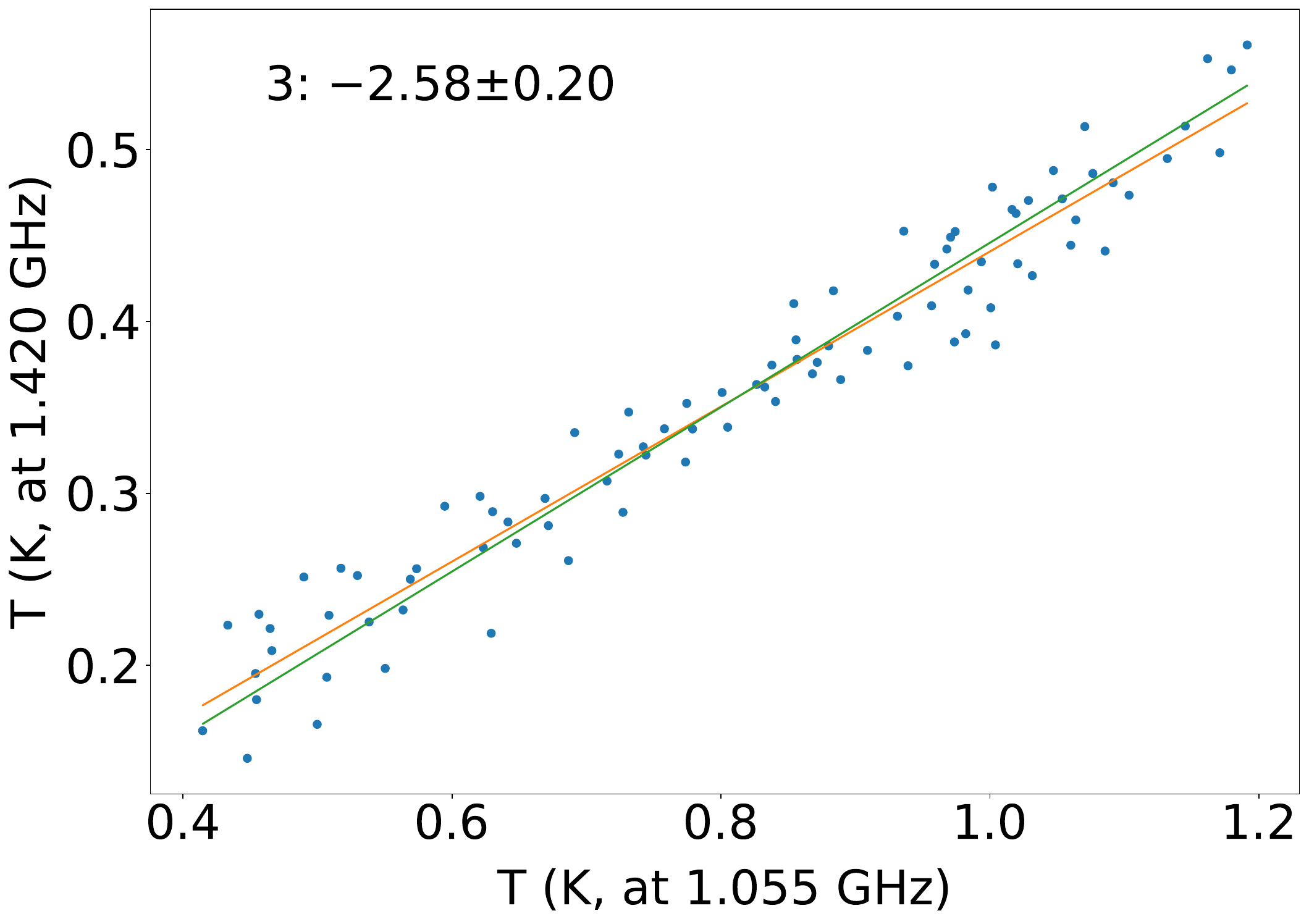}
	\end{subfigure}
	\centering
	\begin{subfigure}{0.49\linewidth}
		\centering
		\includegraphics[width=1\linewidth]{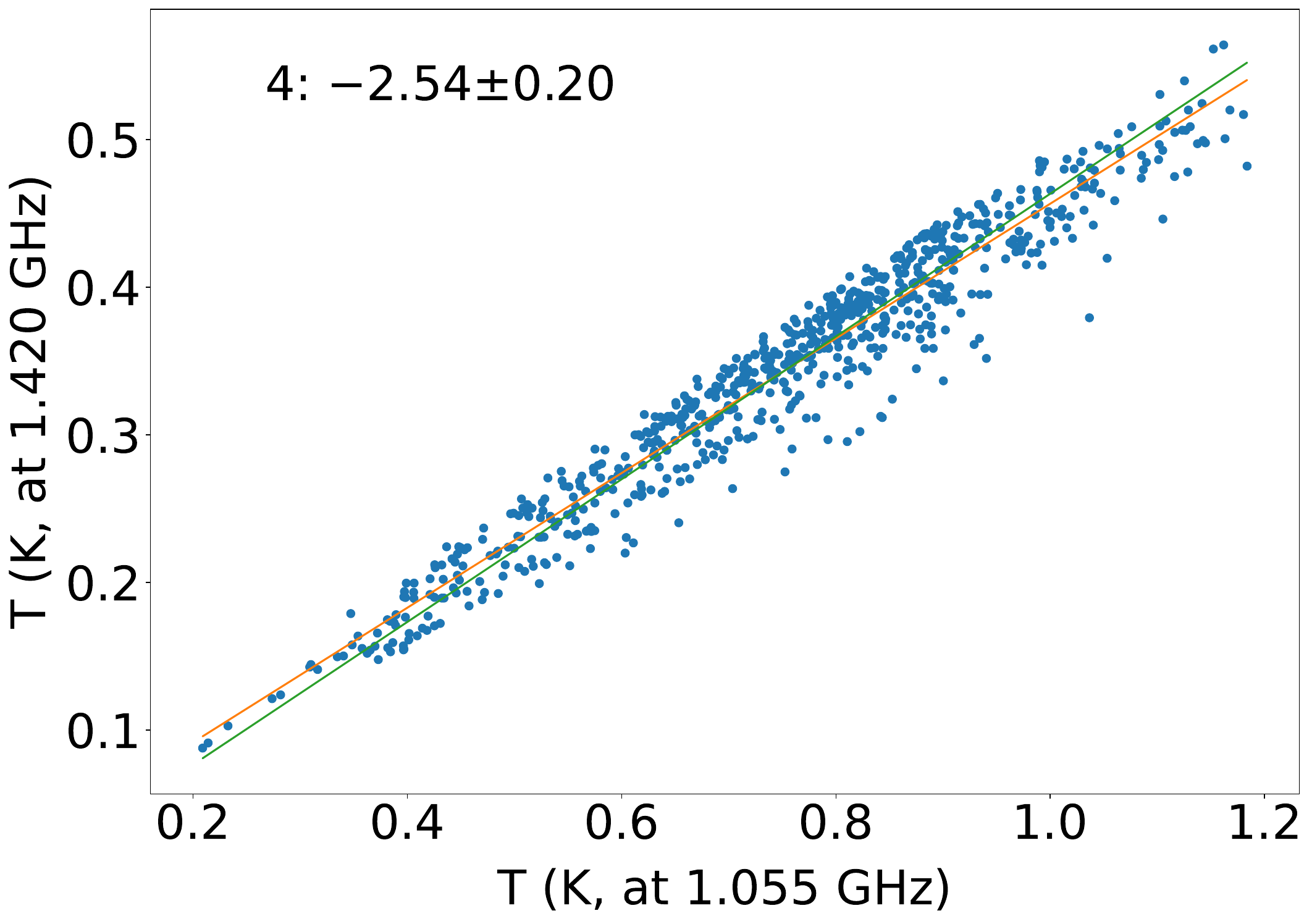}
	\end{subfigure}

	\begin{subfigure}{0.49\linewidth}
		\centering
		\includegraphics[width=1\linewidth]{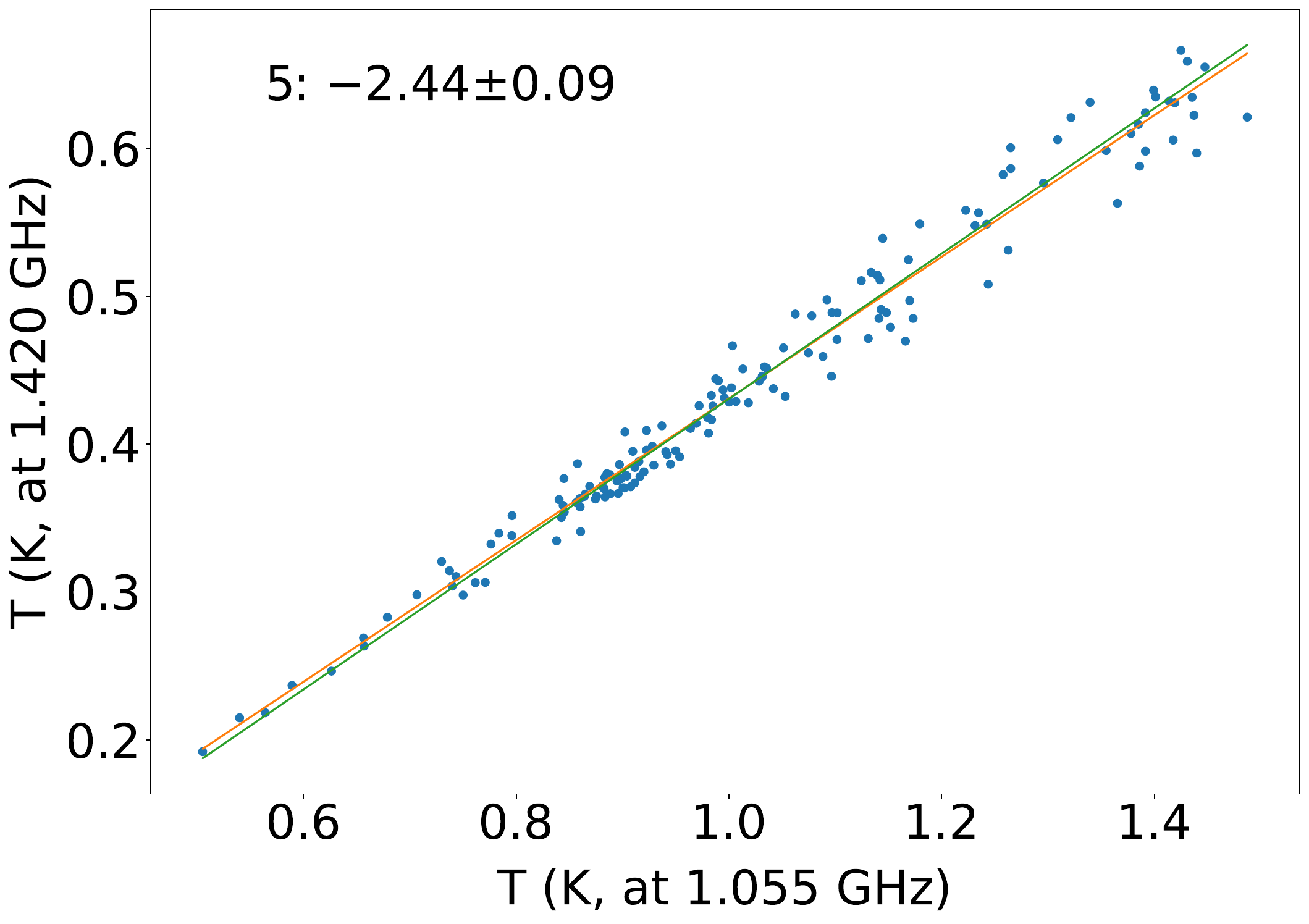}
	\end{subfigure}
	\centering
	\begin{subfigure}{0.49\linewidth}
		\centering
		\includegraphics[width=1\linewidth]{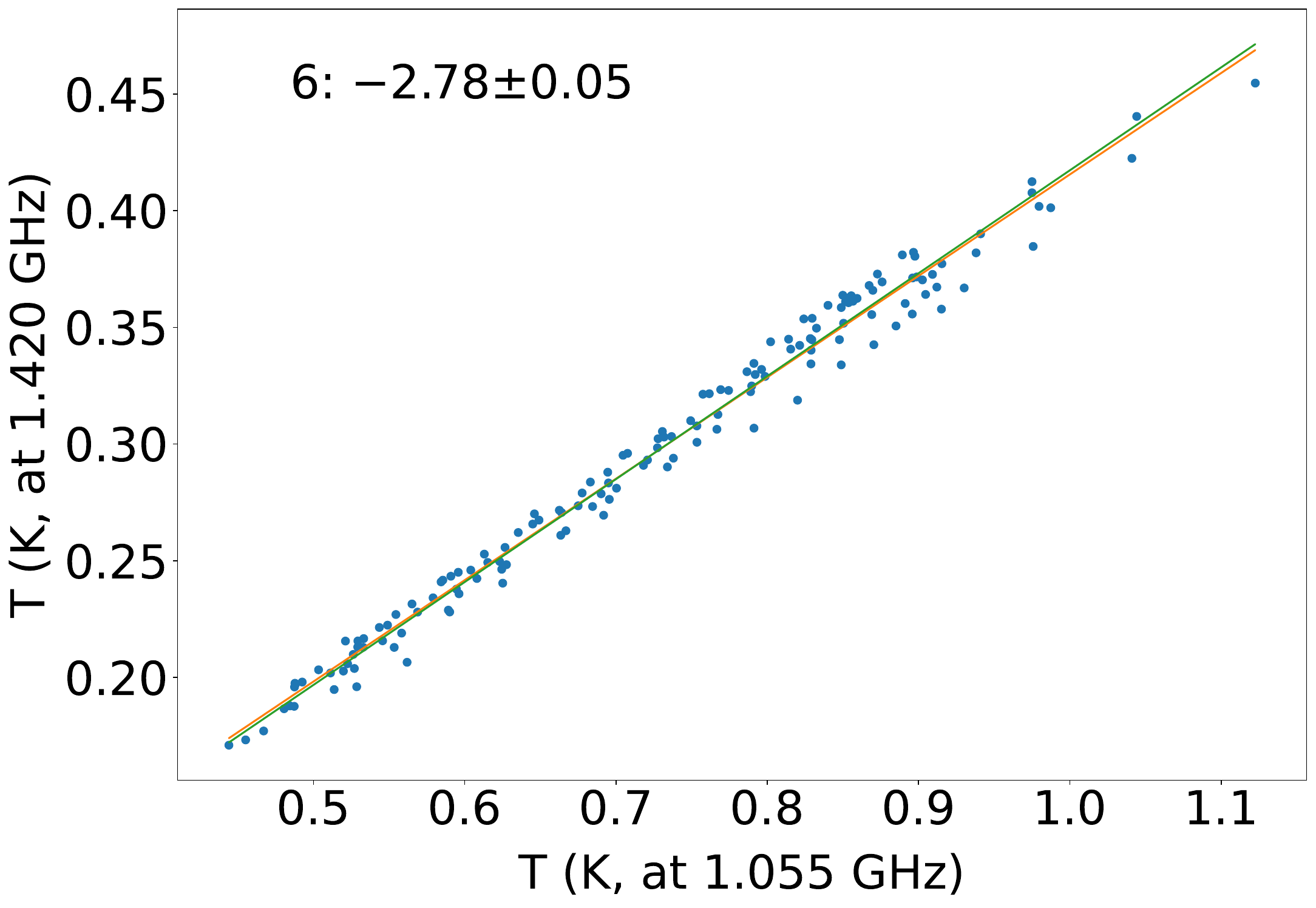}
	\end{subfigure}

	\begin{subfigure}{0.49\linewidth}
		\centering
		\includegraphics[width=1\linewidth]{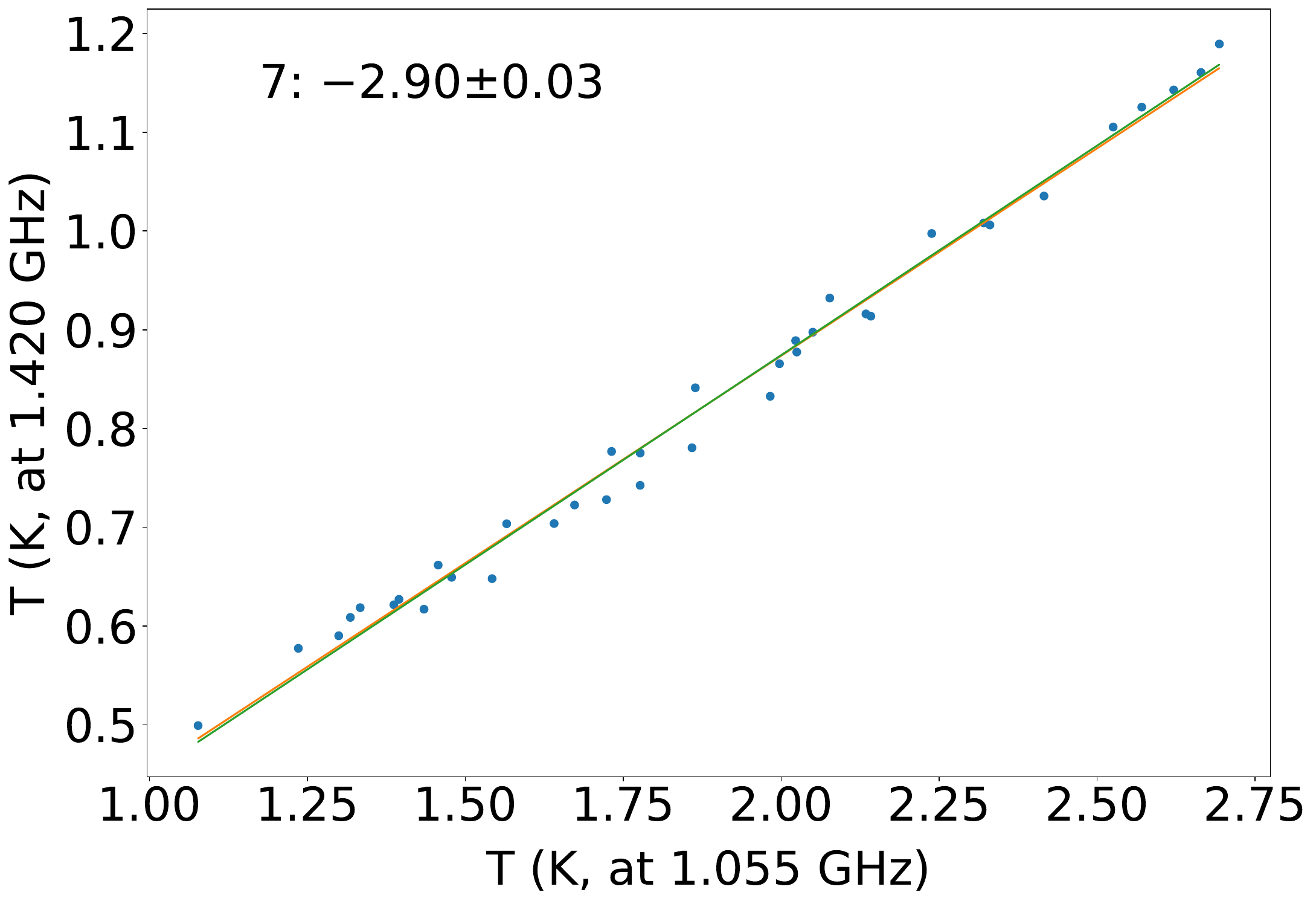}
	\end{subfigure}
        \centering
        \begin{subfigure}{0.49\linewidth}
	\centering
	\includegraphics[width=1\linewidth]{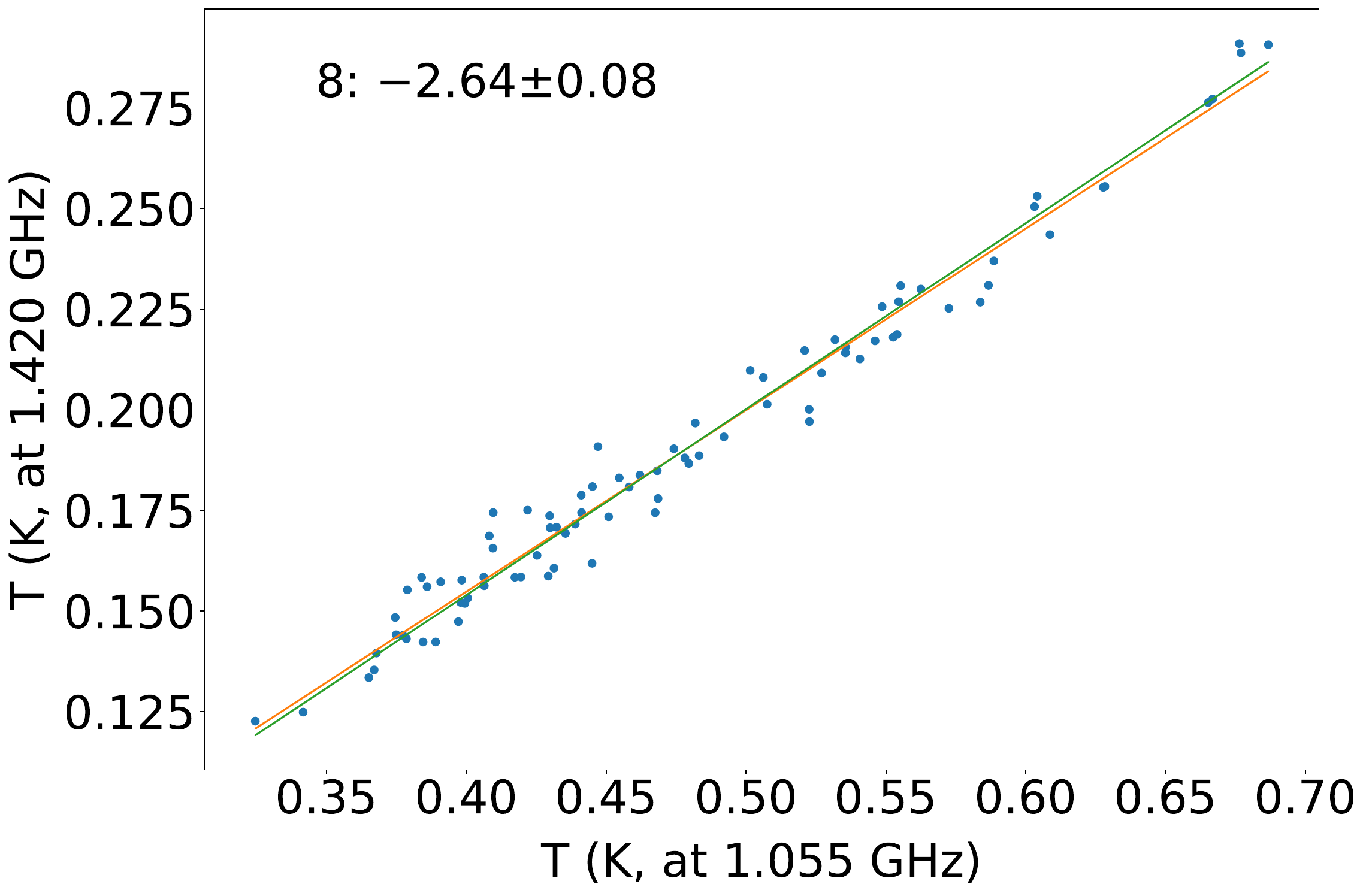}
	\end{subfigure}

	\caption{TT-plots for the areas marked in Fig.~\ref{T-T_picture_mark}. The resulted brightness temperature spectral indices $\beta$ are also shown.}
	\label{T-T picture 7-section}
\end{figure}

We take the median values of the first 10,000 and the last 10,000 channels to calculate the spectral index with TT-plot. The corresponding frequencies are 1.055~GHz and 1.420~GHz, which has the largest frequency separation and hence allows for a more accurate determination of spectral index. The TT-plot results for the 7 regions are shown in Figs.~\ref{T-T_picture_mark} and \ref{T-T picture 7-section}. Note that the brightness temperature spectral index $\beta$, as we obtain, can be connected to the flux density spectral index $\alpha$ as $\alpha=\beta+2$. 

The spectral index for the total intensity is not uniform for M~31. Towards the areas marked as 6, 1, 2 in Fig.~\ref{T-T_picture_mark} where the total intensity emission is weak, the spectra are steep with spectral index $\alpha$ between $-$0.8 and $-0.7$. In contrast, the areas marked as 3, 4, 5 in Fig.~\ref{T-T_picture_mark} where the emission is stronger, the spectra are shallower with spectral index $\alpha$ up to $-0.44\pm0.09$. There seems a tendency of larger intensity with flatter spectrum. The nucleus region marked as 7 has the smallest spectral index of $-0.90\pm0.03$. The area outside M~31 marked as 8, detected because of the high sensitivity of FAST, has a spectral index of $\alpha=-0.64\pm0.08$, and has not been well studied before.

%    \textcolor{blue}{Additionally, we utilized data from the 11~cm \citep{Beck+2020} and FAST telescopes to compute the spectral index of the extended region. the areas marked as 8 in Fig.~\ref{T-T_picture_mark_extend} exhibits a steeper spectral index of $-0.97$ than that observed on the ring. Conversely, a small patch located in the outermost area (the areas marked as 9 in Fig.~\ref{T-T_picture_mark_extend}) has the remarkably flat the spectral index of $-0.05$.}

%\begin{figure}
   %\centering
   %\includegraphics[width=0.98\textwidth, angle=0]{Figure/multiple_ext_region_counter_on_FAST.pdf}
   %\caption{\textcolor{blue}{The spectral index calculation for the extended region is shown in Fig.~\ref{T-T picture extend-section}. The top figure displays the observation data taken at 1.248~GHz by FAST, while the bottom figure the observation at 2.644~GHz conducted by Effelsberg (\citealt{Beck+2020}).}}
   %\label{T-T_picture_mark_extend}
%   \end{figure}

%\begin{figure}[htbp]
	%\centering
	%\begin{subfigure}{0.49\linewidth}
	%	\centering
	%	\includegraphics[width=1\linewidth]{Figure/TT_multiple_ext_region2_1.248 GHz_2.644 GHz.pdf}
	%\end{subfigure}
	%\centering
	%\begin{subfigure}{0.49\linewidth}
	%	\centering
	%	\includegraphics[width=1\linewidth]{Figure/TT_multiple_ext_region3_1.248 GHz_2.644 GHz.pdf}
	%\end{subfigure}

	%\caption{\textcolor{blue}{TT-plots for the extended areas marked in Fig.~\ref{T-T_picture_mark_extend}.}}
	%\label{T-T picture extend-section}
%\end{figure}

\section{Discussion}
\label{sect:discussion}

\subsection{Properties of cosmic ray electrons}

\begin{figure} 
   \centering
   \includegraphics[width=0.98\textwidth, angle=0]{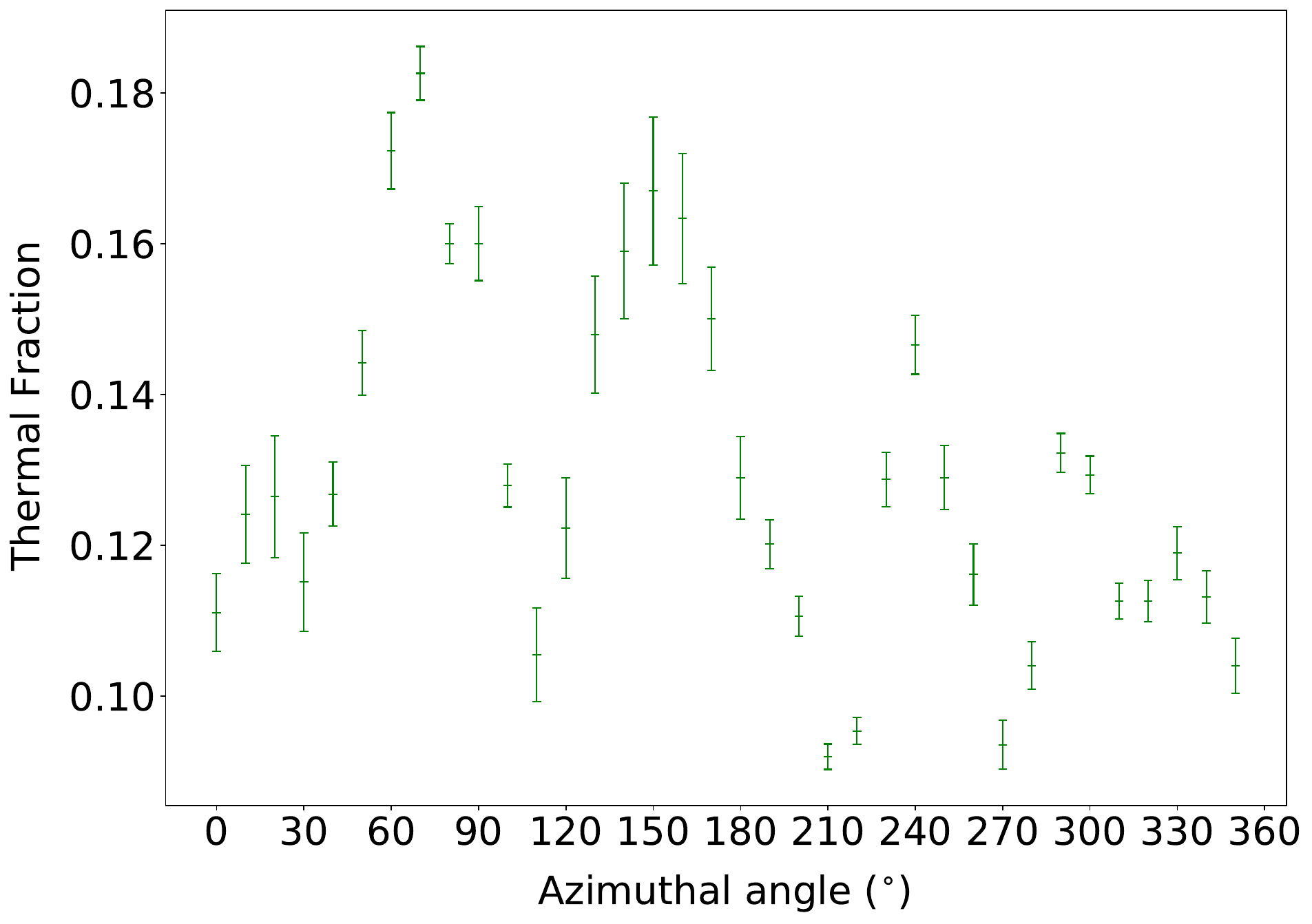}
   \caption{Thermal fraction obtained from \cite{ 2013A&A...557A.129T} versus the azimuthal angle.}
   \label{TF_Azimuthal_angel}
\end{figure}

The total intensity is composed of both thermal emission from free-free radiation and non-thermal emission from synchrotron radiation. We take the images of thermal and non-thermal emission derived by \citet{2013A&A...557A.129T} and obtain the thermal fraction versus the azimuthal angle for the ring of M~31, shown in Fig.~\ref{TF_Azimuthal_angel}. The thermal fraction is up to 20\% for the azimuthal angle between $30\degr$ and $90\degr$, and is about 10\% for the rest. The thermal emission has a flat spectrum with $\alpha_{\rm th}\approx -0.1$ which is much larger than that of non-thermal emission. Therefore the mixture of thermal and non-thermal emission will bring the spectral index of the total emission to a larger value. Typically, for a thermal fraction of 10\% and 20\%, the spectral index of total intensity calculated from the frequency pair of 1.055 and 1.420~GHz is increased by 0.1 and 0.2 in comparison with the spectral index of non-thermal emission $\alpha_{\rm n}$, respectively.

For the emission ring of M~31, the spectra for total intensity towards the north part in Fig.~\ref{tot_intensity} or the regions 3, 4, 5 on the left part in Fig.~\ref{T-T_picture_mark} are shallower than those towards the south part. Taking into account the influence of thermal emission, the spectral index of the non-thermal emission could be estimated by subtracting 0.1 or 0.2 from the total intensity spectral index depending on the thermal fraction. For example, the non-thermal spectral index could reach $\alpha_{\rm n}\approx-0.64$ for region 5 with thermal fraction of about 20\%. This implies that the non-thermal spectral index towards the northern part is larger than that towards the southern part of M~31. 

The asymmetry of the non-thermal spectral index between the northern and southern part of M~31 was also found by \citet{2021A&A...651A..98F} based on the observations at frequencies between 1.5 and 6.6~GHz. The star formation rate is also higher towards the north~\citep{2021A&A...651A..98F}. A higher star formation rate results in a larger number of stars and consequently a larger number of supernovae and supernova remnants that can produce more cosmic ray electrons. Therefore the spectrum is flatter towards the north. Moreover, the presence of higher cosmic ray electron density results in stronger synchrotron radiation, which explains the correlation between large intensity and large spectral index.

For area 8 in Fig.~\ref{T-T_picture_mark}, it cannot be the extension from the ring. Otherwise its spectrum should be steeper because of the aging of cosmic ray electrons as they propagate from the ring. From the Herschel image at 250~$\mu$m~\citep{viaene+14}, there is strong emission corresponding to this area, which seems to be part of a large ring structure surrounding M~31. There could be star forming activities that produce high energy electrons. High resolution observations are needed to detect radio emission related with the infrared emission observed by Herschel.
%the areas marked as 8 in Fig.~\ref{T-T_picture_mark_extend} has a notably steeper spectral index compared to the areas marked as 1 in Fig.~\ref{T-T_picture_mark}. This can be explained by the shorter lifetimes of higher-energy cosmic ray electrons as they propagate from the ring to the outer regions. As a consequence, there is a faster loss of higher frequency cosmic ray electrons, leading to a pronounced steepening of the spectral index~\citep{Beck+2020}. In the regions marked as 9 in Figure \ref{T-T_picture_mark_extend}, characterized by a flat spectral index, higher intensity levels are observed along with accompanying infrared emissions~\citep{Gordon+2006}. This suggests the presence of a amount of dust in this region, resulting in strong thermal emissions and consequently, a flattened spectral index.}

The scale length of the total emission is 3.7~kpc. The thermal emission is highly concentrated in the ring~\citep{2013A&A...557A.129T}, and therefore has little influence on the scale length of non-thermal emission. In practice, we derive a scale length of 3.8~kpc for the non-thermal emission after removing the contribution of thermal emission. This corresponds to an exponential scale length of $\sqrt{2}\times 3.8\approx 5.4$~kpc. Assuming an energy equipartition between magnetic field and cosmic rays, the scale length of the non-thermal emission is equal to that of cosmic ray electrons multiplied by the factor $(3-\alpha_{\rm n})/2\approx 2$~\citep{2013MNRAS.435.1598B}. The scale length of cosmic ray electrons is thus about 2.7~kpc.

The diffusion length of cosmic ray electrons varies with frequency as $\nu^{-\frac{1}{4}}$~\citep{mulcahy+16}. It is therefore expected that the scale length of cosmic ray electrons is about 1.9~kpc at 4.8~GHz, which is consistent with the scale length of about 3.7~kpc for non-thermal emission and hence the scale length of about 1.9~kpc for cosmic ray electrons derived by \citet{Beck+2020} at 4.8~GHz.  This implies that cosmic ray transports primarily by diffusion and the energy loss is not significant at 4.8~GHz.

\subsection{The magnetic field in the ring}

\begin{figure} 
   \centering
   \includegraphics[width=0.98\textwidth,
angle=0]{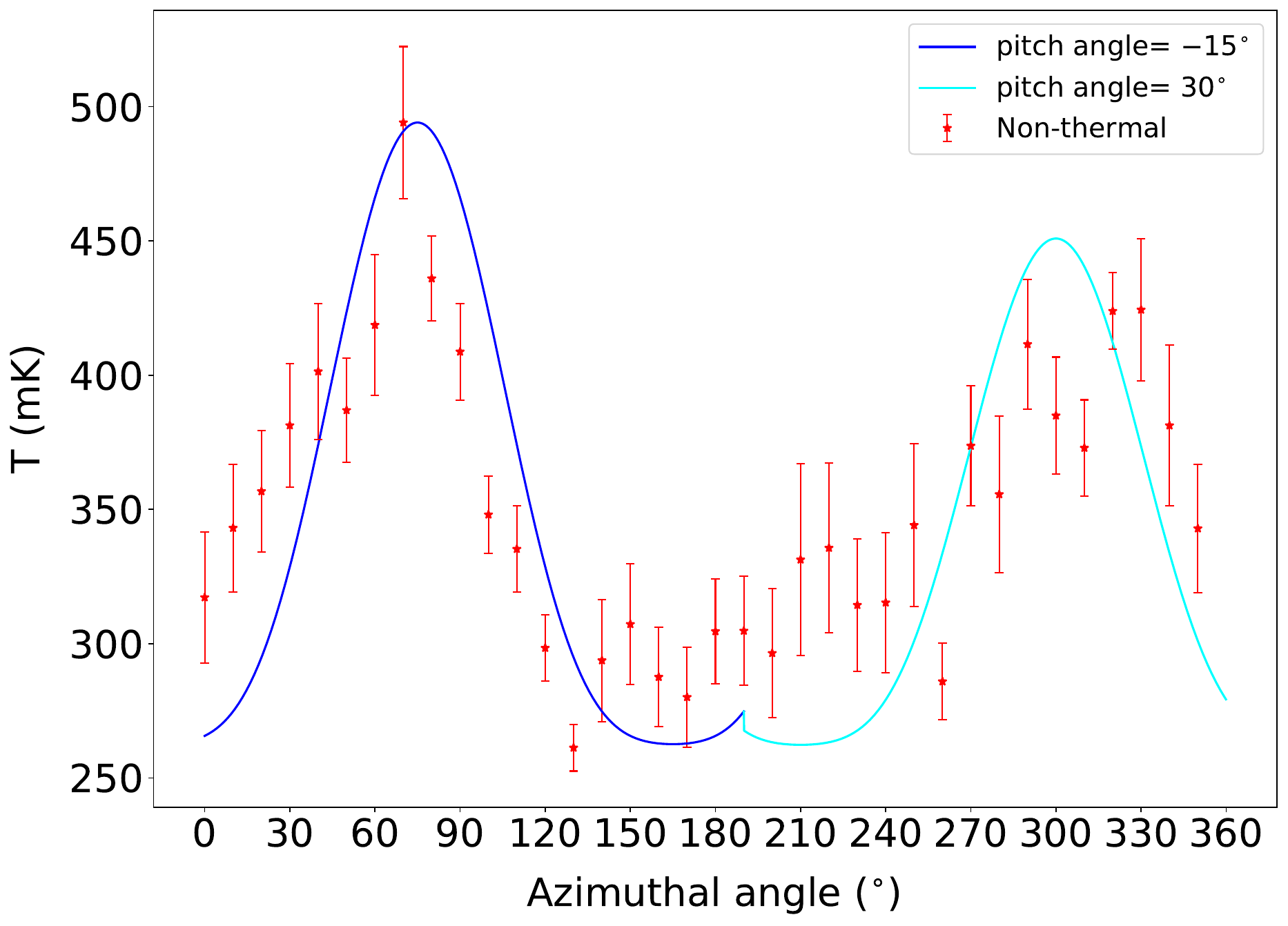}
   \caption{The azimuthal profile of the non-thermal emission within the ring. The lines indicate the profiles expected from an axisymmetric magnetic field with a pitch angle of $-15\degr$ and $30\degr$, respectively.} 
   \label{ASS_Azimuthal_angle_sector10}
\end{figure}

We obtain the azimuthal profile of the non-thermal emission within the bright ring after subtracting the thermal emission~(Fig.~\ref{TF_Azimuthal_angel}) from the total emission~(Fig.~\ref{I_Azimuthal_angle}). The result is shown in Fig.~\ref{ASS_Azimuthal_angle_sector10}. The non-thermal emission is from synchrotron radiation and its profile thus reflects the magnetic field configuration in M~31.

The large-scale regular magnetic field in M~31 mainly follows an axisymmetric pattern~\citep{1998A&A...335.1117H,2004A&A...414...53F}.  In this configuration, the magnetic field perpendicular to the line of sight can be derived as, 
\begin{equation}\label{eq:B}
   B_\perp = B\sqrt{1-\cos^2(\phi-\xi)\sin^2 i}    
\end{equation}
where $\phi$ is the azimuth angle, $\xi$ is the pitch angle, and $i$ is the inclination angle of M~31. The non-thermal intensity ($I$) can be assessed as~\citep{2005AN....326..414B}, 
\begin{equation}\label{eq:I}
    I\propto B_\perp^{3-\alpha_{\rm n}}
\end{equation}
Here the energy equipartition between magnetic field and cosmic rays is assumed.

The random magnetic field has recently been classified into ordered field and isotropic field to interpret the polarization observations of M~31~\citep{Beck+2020}. The ordered field can be generated from shear of differential rotation or compression of spiral arm shocks~\citep{jaffe+19}. However, there are no indications of such large-scale structures inside the ring from radio observations. We therefore only consider the isotropic random magnetic fields. Since the path length along different line of sight towards the ring is similar, the synchrotron emission from the random fields is independent on the azimuthal angle. The azimuthal variation of non-thermal emission is thus caused by the large-scale regular field.

The intensity of synchrotron radiation can be derived using Eqs~\ref{eq:B} and \ref{eq:I}. The inclination angle $i=75\degr$~\citep{2009ApJ...705.1395C} is used. The pitch angle is uncertain. \citet{Berkhuijsen+2003} obtained a pitch angle of $\xi=-13\degr\pm7\degr$ by analyzing polarization observations at 6~cm. \citet{2004A&A...414...53F} found a radial change of the pitch angle from $-17\degr$ to $-8\degr$ based on multi-frequency polarization observations. Whereas, the recent analysis by \citet{Beck+2020} suggested a large variation of the pitch angle versus azimuthal angle for the total magnetic field including both regular and ordered field. 

Assuming a constant pitch angle for M~31, we would expect the separation of the two intensity peaks to be $180\degr$, which is clearly inconsistent with the observed profile shown in Fig.~\ref{ASS_Azimuthal_angle_sector10}. This implies variations of the pitch angle. 
%However, a magnetic field model with a single pitch angle is insufficient to explain the observed profiles which exhibit two distinct peaks in phase. 
%We proposed a magnetic field model that Assumed different pitch angles at different positions along the ring \citep{Beck+2020}. Specifically, we assumed 
We thus propose a simple model of the large scale magnetic field with a pitch angle of $-15\degr$ for the azimuthal angle range of $0\degr$ to $190\degr$, and a pitch angle of $30\degr$ for the azimuthal angle range of $190\degr$ to $360\degr$. Additionally, we impose a factor of 0.95 for the ratio of the two peaks, which we attribute to the variation of cosmic ray electron density.
%reduced the peak between $190\degr$ to $360\degr$ by a scaling factor of 0.95, as the electron density of cosmic rays may vary along the ring. 
The resulted profile is presented in Fig.~\ref{ASS_Azimuthal_angle_sector10}, which reasonably %demonstrating its ability to 
reproduces the observed profile. A more sophisticated model of magnetic field by requires both total intensity and polarization observations at multiple frequencies, and is beyond the scope of this paper.  
%These findings indicate the complex nature of the pitch angle along the ring of M~31. In contrast to previous studies on polarization and Rotation Measure (RM) of M~31 \citep{Berkhuijsen+2003}, we did not identify a large-scale ordered magnetic field in the azimuthal variation of total intensity. To further investigate the precise variation of the pitch angle with azimuth, higher-resolution polarization observations are necessary.}

%\st{We derived the intensity of synchrotron radiation according to Eqs.~\ref{eq:B} and \ref{eq:I}, using the inclination angle $i=75\degr$ \mbox{\citep{2009ApJ...705.1395C}} and the pitch angle of $\xi=-12\degr$ obtained from multi-wavelength radio polarization observations by \citet{2004A&A...414...53F}. It can be clearly seen that the observed profile cannot be reproduced. In order to account for the intensity peak at azimuthal angle of about $40\degr$, the pitch angle is required to be about $-50\degr$. The ASS field with this pitch angle, however, produces intensity largely deviate from the observation at other azimuthal angles. This indicates that the large-scale regular field is more complicated than the ASS field.}

\section{Summary}
\label{sect:Summary}
We obtain the total intensity images from the FAST legacy survey of M~31, covering the frequency range of 1.0 - 1.5~GHz. By averaging over all the frequency channels, we derive the total intensity image at 1.248~GHz, which reveals weak extended emission outside the ring because of the high sensitivity. The image allows us to analyze the propagation of cosmic ray electrons and the magnetic field configuration.

The main conclusions are as follows:
\begin{enumerate}
    \item We derive a scale length of about 2.7~kpc for the cosmic ray electrons by fitting the radial profile of non-thermal emission. In comparision with the scale length at higher frequencies, we find that the cosmic ray electrons propagate mainly through diffusion. 
    \item The spectral index of the total intensity in the ring, calculated from TT-plots, clearly exhibits an azimuthal variation, which can be attributed to the spectral variation of synchrotron radiation. The variation is likely caused by the change of star formation rates along the ring. 
    \item The weak emission detected outside the ring might not be the extended emission from the ring. Instead, it might be produced by star forming activities.
    %but Cosmic ray electrons propagate from the ring to the extended region, resulting in a steeper spectral index. Interestingly, a small area at the edge of the extended region exhibits a remarkably flat spectral index. We assume that this region is the result of significant thermal emissions generated by a dust cloud.}
    \item The azimuthal profile of non-thermal emission can be reproduced with an axisymmetric large-scale magnetic field pattern with pitch angle varying with azimuthal angle, indicating that the magnetic field configuration in the ring is complicated.
    
\end{enumerate}

\normalem
\begin{acknowledgements}
W. Z. and X. S. are supported by the National SKA Program of China (Grant No. 2022SKA0120101). We thank the anonymous referee for the comments that help improve the paper.
\end{acknowledgements}

\bibliographystyle{raa}
\bibliography{bibtex}

\end{document}